\documentclass[]{aa}

\usepackage[draft, layout={inline,nomarginclue}]{fixme}
\fxuseenvlayout{colorsig}
\makeatletter
\renewcommand*\FXLayoutInline[3]{
  {\@fxuseface{inline}\ignorespaces[#3 \fxnotename{#1}: #2]}}
\makeatother

\usepackage{graphicx}
\usepackage{mathtools}
\usepackage{txfonts} 
\usepackage{booktabs}

\begin{document}

   \title{Establishing a relationship between the cosmological 21 cm power spectrum and interferometric closure phases}
   \titlerunning{Relationship between the cosmological 21 cm power spectrum and closure phases}
   \author{P. M. Keller
          \inst{1}
          \and
          B. Nikolic
          \inst{2}
          \and 
          N. Thyagarajan
          \inst{3}
          }
    
   \institute{\inst{1} Leiden Observatory, Leiden University, PO Box 9513, 2300 RA Leiden, The Netherlands \\
              \email{pkeller@strw.leidenuniv.nl}  \\
              \inst{2} Cavendish Astrophysics, University of Cambridge, Cambridge CB3 0HE, UK \\
              \email{bn204@cam.ac.uk}\\
              \inst{3} Space \& Astronomy, Commonwealth Scientific and Industrial Research Organisation (CSIRO), P. O. Box 1130, Bentley, WA 6102, Australia \\
              \email{Nithyanandan.Thyagarajan@csiro.au}
             }

   \date{Received ; accepted}

  \abstract
   {Measurements of the cosmic 21\,cm background need to achieve a high dynamic range to isolate it from bright foreground emissions. Instrumental calibration inaccuracies can compromise the spectral fidelity of the smooth foreground continuum, thereby limiting the dynamic range of the measurement and potentially precluding the detection of the cosmic line signal. In light of this calibration challenge, recent work has proposed using the calibration-independent closure phase to search for the spectral fluctuations of the cosmic 21\,cm background signal. However, so far there has been only a heuristic understanding of the mapping between closure phases and the cosmological power spectrum of the background line signal.}
   {This work aims to establish a more accurate mathematical relationship between closure phase measurements and the cosmological power spectrum of the background line signal.}
   {Building on previous work, we treat the cosmic signal component as a perturbation to the closure phase and use a delay spectrum approach to estimate the power of the perturbing signal. We establish the relationship between this estimate and the cosmological power spectrum using standard Fourier transform techniques, and validate it using simulated observations from the Hydrogen Epoch of Reionization Array (HERA).}
   {We find that, statistically, the power spectrum estimate from closure phases is approximately equal to the true cosmological power spectrum convolved with a foreground-dependent window function, provided that the signal-to-foreground ratio is small. Compared with standard approaches, the foreground dependence of the window function results in an increased amount of mode-mixing and a more pronounced proliferation of foreground power along the line-of-sight dimension of the cylindrical power spectrum. These effects can be mitigated by flagging instances where the window function is broad. Crucial to gaining the necessary sensitivity, this mapping will allow us to average the measurements of closure triads of different shapes based on their imprint in cylindrical Fourier space.}
   {}

   \maketitle

\section{Introduction}

The 21\,cm line of neutral hydrogen (\textsc{H\,i}) is a rich probe of astrophysics and cosmology across a wide range of cosmic epochs \citep[see reviews by][]{pritchard2012, furlanetto2016}. At the redshifts of the dark ages ($z \gtrsim 30$), the \textsc{H\,i} 21\,cm signal is expected to trace the density of baryonic matter and hence will be a pristine probe of early structure formation. Following cosmic dawn ($z \sim 30$), the thermal state of hydrogen coupled to the photons emitted by astrophysical sources, which culminated in the reionisation of the intergalactic medium (IGM) expected to have occurred at $z \sim 6-12$. The \textsc{H\,i} 21\,cm signal from cosmic dawn and the epoch of cosmic reionisation (EoR) should be rich in information about the first generation of stars and black holes. In the post-reionisation Universe, neutral hydrogen resides predominantly in galaxies. The cosmic \textsc{H\,i} 21\,cm signal at these epochs can thus be used to detect Baryon acoustic oscillations (BAO), which serve as a standard ruler for measuring the expansion history of the Universe. Such measurements are of particular interest at the redshifts where cosmic expansion started to be increasingly influenced by dark energy ($z \lesssim 3$). 

Several current and future interferometric experiments aim to detect the spatial fluctuations of the cosmic \textsc{H\,i} 21\,cm signal. Low-frequency experiments ($\sim 30-200$\,MHz) targeting cosmic dawn and the EoR include HERA \citep[][]{hera2017}, the MWA \citep[][]{mwa}, LOFAR \citep[][]{lofar}, NenuFAR \citep[][]{nenufar}, the OVRO-LWA \citep[][]{lwa}, the GMRT EoR experiment \citep[][]{gmrt}, 21CMA \citep[][]{21cma} and the SKA-low \citep[][]{skaEoR}. At higher frequencies ($\sim 200-1000$\,MHz), CHIME \citep[][]{chime}, HIRAX \citep[][]{hirax}, TIANLAI \citep[][]{tianlai}, PUMA \citep[][]{puma}, and SKA-mid \citep[][]{ska_im} aim for BOA detections at $z \lesssim 3$. 

The common challenge shared by these experiments are the very bright foreground emissions which overwhelm the faint cosmic \textsc{H\,i} 21\,cm signal by many orders of magnitude. The line-of-sight fluctuations of the cosmic line signal offer a means by which it can, in principle, be separated from the spectrally smooth continuum foregrounds \citep{datta, dspec,Thyagarajan+2013, eorwindowmath}. This separation requires spectral calibration inaccuracies to be smaller than the spectral fluctuations of the cosmic line signal. Instrumental systematics and modelling uncertainties can prevent conventional calibration methods from achieving the necessary spectral fidelity, thereby ruling out a detection of the cosmic line signal \citep{barry2016calibration, ewall2017calibration, byrne2019limitations, byrne2021unified}.

As an alternative to standard analysis approaches, \cite{closure1} propose using interferometric closure phases to detect the spectral fluctuations of the cosmic \textsc{H\,i} 21\,cm signal. Closure phases are independent of antenna-based, direction-independent calibration, making them insensitive to a wide range of instrumental systematics \citep{TMS2017, Thyagarajan+2022, samuel+2022}. By evading the conventional calibration step, the closure phase approach simplifies the data processing and reduces the chance of introducing processing-based systematics. 

\cite{closuremaths} motivate a closure phase based estimator of the power spectrum of the cosmic line signal. Variations of this estimator were used to set approximate upper limits on the EoR \textsc{H\,i} 21\,cm power spectrum with data from HERA \citep[][]{closurelimits, Keller2023}, and the MWA \citep[][]{Tiwari2024}. The estimators used in these papers differ mainly by the heuristic scaling used to map phase fluctuations to fluctuations in brightness temperature. The lack of consensus on this scaling reveals a hitherto limited understanding of the relationship between closure phases and the cosmic \textsc{H\,i} 21\,cm power spectrum. In this paper, we aim to derive a more accurate description of this relationship from which the correct scaling of the power spectrum estimator arises naturally.

The paper is structured as follows. Section~\ref{sec:maths} derives a mapping between closure phases and the cosmic power spectrum in the limit of small fluctuations. Section~\ref{sec:validation} describes the methods and simulations used to validate the power spectrum estimator. The results of this validation work are presented and discussed in Section~\ref{sec:results}. Section~\ref{sec:conclusion} concludes the paper.

\section{Mathematical foundations}
\label{sec:maths}
The fundamental quantity of radio interferometric measurements is the complex visibility, formed by correlating the electric signals between pairs of antennas. Under certain commonly assumed conditions, the visibility can be shown to represent a Fourier component of the orthographically projected apparent sky intensity \citep[e.g.][]{TMS2017}:
\begin{align}
    \label{eq:vis}
    V(\mathbf{b}_\nu , \nu) &= \smashoperator{\iint\limits_{-\infty}^{+\infty}} dldm A(\hat{\mathbf{s}}, \nu) I(\hat{\mathbf{s}}, \nu)  e^{-2\pi i \hat{\mathbf{s}} \cdot \mathbf{b}_\nu}\\
    \label{eq:vis2}
    &= \widetilde{A}(\mathbf{b}_\nu , \nu) * \widetilde{I}(\mathbf{b}_\nu , \nu) 
\end{align}
where $\nu$ is the frequency, $\mathbf{b}_\nu$ is the antenna baseline vector in units of wavelengths, $\hat{\mathbf{s}} = (l, m)$ are the orthographic sky coordinates, $I(\hat{\mathbf{s}}, \nu)$ is the specific intensity, and $A(\hat{\mathbf{s}}, \nu)$ describes the antenna primary beam and other direction-dependent effects, if applicable. The second line uses the convolution theorem, where the convolution is denoted by an asterisk and the tildes indicate Fourier-transformed quantities. At the frequencies of interest to 21\,cm cosmology, the ionosphere leaves a strong directional and time-dependent imprint on the apparent sky. However, at baselines shorter than a few kilometres all antennas essentially see the same part of the ionosphere, so that the ionospheric phase shifts cancel in the signal correlation process. In this work, we consider only baselines shorter than 200\,m for which these assumptions are reasonable \citep{Gasperin2018}. For notational convenience, we omit measurement noise throughout this work.

\subsection{Closure phase}
Actual interferometric measurements are corrupted by various effects along the signal propagation path. The dominant corruptions can usually be represented as antenna-based, direction-independent complex gain factors. The visibility, $V_{pq}$, measured by antennas $p$ and $q$, can thus be written as
\begin{equation}
    V_{pq} = g_p^{} g_q^* \widehat{V}_{pq} = \left|g_p\right| \left|g_q\right| \left|\widehat{V}_{pq}\right| e^{i(\phi_{pq} + \varphi_p - \varphi_q)},
\end{equation}
where $\widehat{V}_{pq}$ is the uncorrupted visibility, $\phi_{pq} = \arg(\widehat{V}_{pq})$ is the sky-based phase term, $g_p$ are the gains and $\phi_p = \arg(g_p)$ their respective phases, and the asterisk denotes complex conjugation. The frequency and time dependence of the gains and the visibilities are left implicit in this equation. 

Consider a triad formed by antennas $m$, $p$ and $q$. The closure phase is obtained by summing the visibility phases measured on this triad in a closed loop:
\begin{align}
    \nonumber \phi_{mpq} &= \arg(V_{mp}) + \arg(V_{pq}) + \arg(V_{qm}) \\ \nonumber
    &= \arg(\widehat{V}_{mp}) + \arg(\widehat{V}_{pq}) + \arg(\widehat{V}_{qm}) \\
    &= \phi_{mp} + \phi_{pq} + \phi_{qm},
\end{align}
where the gain phases cancel out in the second line. The closure phase is therefore free of antenna-based, direction-independent corruptions and depends only on sky-based terms \citep{closurephase,Thyagarajan+Carilli2022}. 

\begin{figure}
    \centering
    \includegraphics[width=0.8\linewidth]{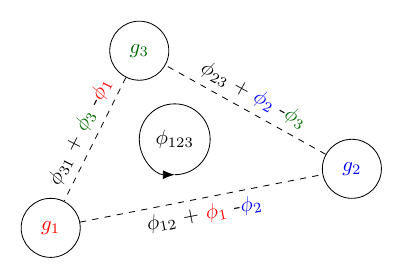}
    \caption[Closure phase illustration]{Illustration of the closure phase concept. The true visibility phases, $\phi_{pq}$, are corrupted by direction-independent antenna-based gains, $\phi_{p}$. Summing the corrupted phases in a loop eliminates the gain phases. The resulting quantity, $\phi_{123}$, is the closure phase.}
    \label{fig:cp}
\end{figure}

\subsection{Perturbation theory}
\label{sec:pert}
The effect of the cosmic line signal on the closure phase is best described using perturbation theory. \cite{closuremaths} provide an extensive discussion on closure phase perturbations. For completeness, we repeat the main points relevant to this paper.

\begin{figure}
    \centering
    \includegraphics[width=0.8\linewidth]{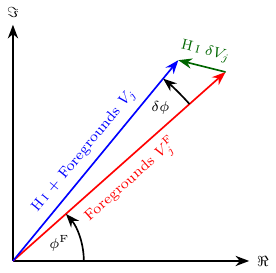}
    \caption[Illustration of a visibility phase perturbation]{Exaggerated illustration of the visibility phase perturbations due to the weak cosmic signal $\delta V_j$ (green arrow). The horizontal and vertical axes represent the real and complex axes of the complex plane, respectively. The red arrow represents the foreground visibility, $V_j^\mathrm{F}$, and the blue arrow represents the sum of the foreground and cosmic \textsc{H\,i} 21\,cm visibilities, $V_j = V_j^\mathrm{F} + \delta V_j$. The resulting phase perturbation is $\delta \phi$.}
    \label{fig:cp_pert}
\end{figure}

The uncorrupted visibilities are typically dominated by bright continuum foreground sources. Hence, we can treat the weak cosmic signal as a perturbation to the uncorrupted foreground signal. The measured visibility can then be written as 
\begin{equation}
    V_{j} = V^\mathrm{F}_{j} + \delta V_{j},
\end{equation}
where the subscript $j$ indexes the baseline, the superscript F denotes the foregrounds and $\delta V_{j}$ is the perturbation such that $|\delta V_{j}/V^\mathrm{F}_{j}|\ll 1$ (see Figure~\ref{fig:cp_pert}). The change from indexing quantities by antenna pairs to indexing them by their baseline is notationally more convenient for this section.

Remembering that $\phi_{j}=\tan^{-1}(\Im\{V_{j}\}/\Re\{V_{j}\})$ and expanding to first order in $\delta V_{j}$, one finds
\begin{align}
    \label{eq:phase_taylor}
    \phi_{j} &\approx \phi^\mathrm{F}_{j}+\Im\left\{\frac{\delta V_{j}}{V_{j}^\mathrm{F}}\right\},
\end{align}
where $\phi_{j}^\mathrm{F}$ is the unperturbed visibility phase. The perturbed closure phase $\phi_\triangledown$ is then simply the sum of three perturbed visibility phases:
\begin{equation}
    \label{eq:cp_pert}
    \phi_\triangledown \approx \phi_\triangledown^\mathrm{F} + \sum_{j=1}^{3} \Im\left\{\frac{\delta V_{j}}{V_{j}^\mathrm{F}}\right\} = \phi_\triangledown^\mathrm{F} + \frac{1}{2i}\sum_{j=1}^{3} \left[\frac{\delta V_{j}}{V_{j}^\mathrm{F}} - \frac{\delta V_{j}^*}{V_{j}^{\mathrm{F}*}}\right],
\end{equation}
where $\phi_\triangledown^\mathrm{F}$ is the unperturbed closure phase and the summed-over baselines form a closed loop in real space. The right side follows from the complex identity $\Im \{z\} = (z - z^*)/2i$. \cite{closuremaths} find that the fractional inaccuracy of this fist-order approximation is smaller than $\sim 10^{-4}$ for the expected strength of the EoR \textsc{H\,i} 21\,cm signal.

The equations above establish that the strength of phase perturbations is proportional to the complex ratio of the perturbing visibilities to the foreground visibilities. Since it is the imaginary parts of these ratios that are relevant, only the perturbing components that are perpendicular to the foreground visibilities in the complex plane induce phase perturbations. \cite{closuremaths} therefore interpret closure phase perturbations as a measure of the relative dissimilarity between the cosmic signal and the foregrounds. Furthermore, they point out that, statistically, only half of the fluctuations of the cosmic signal are recovered in phase, with the other half contained in the amplitudes. As we work towards an estimator of the cosmological power spectrum, this factor of two appears naturally in the power spectrum scaling. 

\subsection{Closure phase delay spectrum}
\label{sec:cp_pspec}
The delay spectrum approach is a commonly used method of 21\,cm cosmology data analysis, which allows one to access the line-of-sight fluctuations of the cosmic line signal while isolating the smooth continuum emissions of bright foregrounds \citep{datta, dspec}. As demonstrated by \cite{closure1}, a similar approach can be used to recover the cosmic signal from the closure phase. In analogy to the visibility delay spectrum, \cite{closure1} define the closure phase delay spectrum at a given redshift, $z$, as a windowed Fourier transform across frequency:
\begin{equation}
    \label{eq:closure_dspec}
    \widetilde{\Psi}_\triangledown(\tau; \nu_0) = \smashoperator{\int\limits_{-\infty}^{+\infty}} d\nu \, W(\nu-\nu_0) e^{i\phi_\triangledown(\nu)} e^{2\pi i \tau \nu},
\end{equation}
where $W(\nu)$ is a spectral tapering function and $\nu_0=\nu_\textsc{H\,I}/(1+z)$ is the centre frequency of the power spectrum estimation. The delay transform in this equation is performed on the complex exponential of the closure phase rather than on the closure phase directly to avoid discontinuities arising from the phase branch cuts, i.e. at $\pm \pi$ if the phase is defined on $(-\pi, \pi]$. In the presence of a small closure phase perturbation, $\delta \phi_\triangledown$, we can approximate the complex exponential of the closure phase to first order as 
\begin{equation}
    \label{eq:cp_exp}
    e^{i(\phi_\triangledown^\mathrm{F}(\nu) + \delta \phi_\triangledown(\nu))} \approx e^{i\phi_\triangledown^\mathrm{F}(\nu)}\left(1 + i\delta \phi_\triangledown(\nu)\right).
\end{equation}
Inserting this along with Equation~\ref{eq:cp_pert} into the expression for the delay spectrum (Equation~\ref{eq:closure_dspec}) and using the convolution theorem, we find \citep[cf.][]{closuremaths} 

\begin{multline}
    \label{eq:closure_dspec2}
    \widetilde{\Psi}_\triangledown(\tau) = \widetilde{\Psi}_\triangledown^\mathrm{F}(\tau)
    + \frac{1}{2}\sum_{j=1}^{3}\left(\widetilde{\Xi}_j(\tau) * \widetilde{\delta V_{j}}(\tau) - \widetilde{\Xi}_j^*(-\tau) * \widetilde{\delta V_{j}^*}(-\tau)\right),
\end{multline}
where
\begin{align}
    \widetilde{\Psi}_\triangledown^\mathrm{F}(\tau) &= \smashoperator{\int\limits_{-\infty}^{+\infty}} d\nu \, W(\nu-\nu_0) e^{i\phi_\triangledown^\mathrm{F}(\nu)} e^{2\pi i \tau \nu}, \\
    \label{eq:xi}
    \widetilde{\Xi}_j(\tau) &= \smashoperator{\int\limits_{-\infty}^{+\infty}} d\nu \, \frac{W(\nu-\nu_0)}{V_j^\mathrm{F}(\nu)} e^{i\phi_\triangledown^\mathrm{F}(\nu)} e^{2\pi i \tau \nu}, \\
    \widetilde{\delta V_{j}}(\tau) &= \smashoperator{\int\limits_{v_\mathrm{min}}^{v_\mathrm{max}}} d\nu \, \delta V_j(\nu) e^{2\pi i \tau \nu}
    \label{eq:dV}
\end{align}
describe a set of Fourier transforms of the various frequency dependent terms in Equations~\ref{eq:cp_pert}$-$\ref{eq:cp_exp}. In Equation~\ref{eq:closure_dspec2}, the superscript asterisk denotes complex conjugation and the regular asterisk denotes a convolution. In the last equation, the integration limits define the frequency range, $(v_\mathrm{min}$, $v_\mathrm{max})$, in which $W(\nu-\nu_0)$ is non-zero.

\subsection{Estimating the cosmological power spectrum}
The simplest statistic to describe the fluctuations of the cosmological background signal is the power spectrum, $P(\mathbf{k})$, defined through the relation
\begin{equation}
    \label{eq:ps}
    \left<\widetilde{\delta I}(\mathbf{k}) \widetilde{\delta I}^*(\mathbf{k}^\prime)\right> = \left(\frac{2 k_\mathrm{B}}{\lambda^2}\right)^2(2\pi)^3 \delta(\mathbf{k} - \mathbf{k}^\prime) P(\mathbf{k}),
\end{equation}
where the angular brackets denote an ensemble average, $\mathbf{k}$ and $\mathbf{k^\prime}$ are comoving wave numbers, $\widetilde{\delta I}(\mathbf{k})$ is the Fourier transform of the specific intensity of the cosmic background signal, $k_\mathrm{B}$ is Boltzmann's Constant, $\lambda = c / \nu$ is the wavelength, and $\delta(\mathbf{k})$ is a Dirac delta function. The first bracket on the right side of Equation~\ref{eq:ps} is the Rayleigh-Jeans factor, which converts specific intensities to brightness temperatures. 

The following paragraphs derive a relationship between the closure phase delay spectrum and the cosmological power spectrum for arbitrary triad shapes. This can be achieved by computing the ensemble average\footnote{The ensemble average is in practice replaced with an average over different realisations of the sky signal, i.e. over observations of independent portions of the sky, and different triad shapes. The resulting estimate has an associated uncertainty due to cosmic variance.} of the squared closure phase delay spectrum. In doing so, there are various cross-correlations among the terms in Equation~\ref{eq:closure_dspec2}. Assuming that the cosmic line signal is statistically isotropic and uncorrelated with the foregrounds, the only non-zero correlations are of the form 
\begin{multline}
    \label{eq:cp_dspec_corr}
    C(\mathbf{b}, \mathbf{b}^\prime, \tau, \tau^\prime) = \left<\left(\widetilde{\Xi}_j^{}(\tau) * \widetilde{\delta V_{j}^{}}(\tau)\right) \left(\widetilde{\Xi}_{j^\prime}^*(\tau^\prime) * \widetilde{\delta V_{j^\prime}^*}(\tau^\prime)\right)\right> \\
    = \smashoperator{\iint\limits_{-\infty}^{+\infty}} d\eta d\eta' \, \left<\widetilde{\Xi}_j^{}(\tau - \eta) \widetilde{\Xi}_{j^\prime}^*(\tau^\prime - \eta^\prime)\right> \left<\widetilde{\delta V_{j}^{}}(\eta)\widetilde{\delta V_{j^\prime}^*}(\eta^\prime)\right>.
\end{multline}
The second ensemble average in the integrand, i.e. the correlation of the visibility perturbations, can be approximated as
\begin{multline}
    \label{eq:dvis_corr}
    \left<\widetilde{\delta V_{j}^{}}(\tau)\widetilde{\delta V_{j^\prime}^*}(\tau)\right> \approx \smashoperator{\iint\limits_{-\infty}^{+\infty}} d^2u \, d^2u^\prime \, \widetilde{A}(\mathbf{b} - \mathbf{u}) \widetilde{A}^*(\mathbf{b}^\prime - \mathbf{u}^\prime) \\
    \times \left< \widetilde{\delta I}(\mathbf{u} , \tau) \widetilde{\delta I}^*(\mathbf{u}^\prime, \tau) \right>,
\end{multline}
where we used Equation~\ref{eq:vis2}, and implicitly evaluated the effective baseline vector and the antenna beams at $\nu_0$ by making the assumption that they are constant across the bandwidth probed by $W(\nu - \nu_0)$. This is usually a good assumption, as the bandwidth is typically chosen to be a few per cent of the centre frequency to limit the cosmic evolution within the band \citep[see][]{datta2014, eorwindowmath}. 

The observational quantities $\tau$ and $\mathbf{b}$ can be mapped to the cosmological line-of-sight and transverse wave numbers at redshift $z$ via the relations $k_\parallel = 2\pi \tau /X$ and $\mathbf{k}_\perp = 2\pi \mathbf{b_{\nu}} / Y$, where \citep[e.g.][]{morales2004, dspec, eorwindowmath}
\begin{equation}
\label{eq:k}
      X = \frac{c (1 + z)^2}{\nu_\textsc{H\,i} H(z)} \qquad \textrm{and} \qquad Y = D_c(z),
\end{equation}
respectively, $H(z) = H_0 \sqrt{\Omega_\mathrm{M} (1+z)^3 + \Omega_\mathrm{K} (1+z)^2 + \Omega_\Lambda}$ is the Hubble parameter expressed in standard terms of cosmology, $\nu_\textsc{H\,i}$ is the rest frequency of the \textsc{H\,i} 21\,cm line, $c$ is the speed of light, and
\begin{equation}
    D_c(z) = \int_{0}^{z}dz\frac{c}{H(z)} 
\end{equation}
is the comoving line-of-sight distance to the point of the \textsc{H\,i} 21\,cm emission at the frequency $\nu_0$ \citep{cosmo_distances}.

Using the above relations together with the definition of the power spectrum (Equation~\ref{eq:ps}), we can express Equation~\ref{eq:dvis_corr} as \footnote{The $(2\pi)^3$ arises from the fact that $\widetilde{\delta I}(\mathbf{k})$ and $\widetilde{\delta I}(\mathbf{b}, \tau)$ use different Fourier conventions \citep[see][]{liu2020data}.}
\begin{multline}
    \label{eq:dvis_corr_2}
    \left<\widetilde{\delta V_{j}^{}}(\tau)\widetilde{\delta V_{j^\prime}^*}(\tau^\prime)\right> \approx \left(\frac{2k_\mathrm{B}}{\lambda^2}\right)^2\frac{1}{(2\pi)^3} \smashoperator{\int\limits_{-\infty}^{+\infty}} d^2k_\perp \,  \delta\left(k_\parallel - k_\parallel^\prime\right) P\left(\mathbf{k}_\perp, k_\parallel\right) \\
    \times \widetilde{A}\left(\mathbf{b} - \frac{Y \mathbf{k_\perp}}{2\pi}\right) \widetilde{A}^*\left(\mathbf{b}^\prime - \frac{Y \mathbf{k_\perp}}{2\pi}\right).
\end{multline}
Inserting this into Equation~\ref{eq:cp_dspec_corr} gives
\begin{align}
    \label{eq:cp_dspec_corr_2}
    \nonumber C(\mathbf{b}, \mathbf{b}^\prime, \tau, \tau^\prime) &= \left(\frac{2k_\mathrm{B}}{\lambda^2}\right)^2\frac{1}{(2\pi)^3} \smashoperator{\iint\limits_{-\infty}^{+\infty}} d^2k_\perp dk_\parallel \, P(\mathbf{k}_\perp, k_\parallel) \\ \nonumber
    &\times \left<\widetilde{\Xi}_j^{}\left(\tau - \frac{X k_\parallel}{2\pi}\right) \widetilde{\Xi}_j^{*}\left(\tau^\prime - \frac{X k_\parallel}{2\pi}\right)\right> \\ 
    &\times \widetilde{A}\left(\mathbf{b} - \frac{Y \mathbf{k_\perp}}{2\pi}\right) \widetilde{A}^*\left(\mathbf{b}^\prime - \frac{Y \mathbf{k_\perp}}{2\pi}\right).
\end{align}
The second line of this equation is close to zero if $|\tau - \tau^\prime| \Delta \nu > 1$, where $\Delta \nu$ is the width of the effective spectral window function, $W(\nu - \nu_0) / V_j^\mathrm{F}(\nu)$. Here, we are primarily interested in the case where $\tau^\prime = -\tau$ (cf. Equation~\ref{eq:closure_dspec2}) for which $|C(\mathbf{b}, \mathbf{b}^\prime, \tau, -\tau)| \approx 0$ at $\tau$-modes that are not contaminated by foreground power. Similarly, the correlation scale of the terms in the third line of Equation~\ref{eq:cp_dspec_corr_2} depends mainly on the width of the primary beam and the difference between the baseline vectors, $\Delta b = |\mathbf{b} - \mathbf{b}^\prime| c / \nu$. The primary beam width is $\Delta \theta \approx c / (D\nu)$, where $D$ is the diameter of an array element, and therefore the width of $\widetilde{A}(\mathbf{b})$ is approximately $\Delta \theta^{-1} \approx D\nu / c$. The correlation in Equation~\ref{eq:cp_dspec_corr} is therefore small if $\Delta b / D > 1$ \citep{Bharadwaj2001} and in the case of nonidentical baselines it is only relevant for the shortest baselines of very compact arrays such as HERA (to be discussed in Section~\ref{sec:c-terms}).

The following discussion uses the approximation $|C(\mathbf{b}, \mathbf{b}^\prime, \tau, \tau)| \approx 0$ for $\mathbf{b} \neq \mathbf{b}^\prime$, allowing us to write the closure phase delay power spectrum of the perturbing signal as
\begin{align}
    \nonumber \left<\left|\widetilde{\Psi}_\triangledown^\mathrm{H}(\tau)\right|^2\right> &= \frac{1}{4} \sum_{j=1}^{3} q_j \left[ C(\mathbf{b}_j, \mathbf{b}_j, \tau, \tau) + C(\mathbf{b}_j, \mathbf{b}_j, -\tau, -\tau) \right] \\ \nonumber 
    \label{eq:cp_dspec_corr_3}
    &= \frac{1}{2}\left(\frac{2k_\mathrm{B}}{\lambda^2}\right)^2\frac{1}{(2\pi)^3} \smashoperator{\int\limits_{-\infty}^{+\infty}}  d^2k_\perp dk_\parallel \, W_\triangledown(\mathbf{k_\perp}, k_\parallel; \tau) P(k),
\end{align}
where
\begin{equation}
    \label{eq:w_cyl}
    W_\triangledown(\mathbf{k_\perp}, k_\parallel; \tau) = \sum_{j=1}^{3} q_j
    \left<\left|\widetilde{\Xi}_j\left(\tau - \frac{X k_\parallel}{2\pi}\right)\widetilde{A}\left(\mathbf{b}_{j} - \frac{Y \mathbf{k}_\perp}{2\pi}\right)\right|^2\right>
\end{equation}
is the window function, $\widetilde{\Psi}_\triangledown^\mathrm{H}(\tau) = \widetilde{\Psi}_\triangledown(\tau) - \widetilde{\Psi}_\triangledown^\mathrm{F}(\tau)$ is the closure phase delay spectrum of the perturbing signal, $q_j=2$ if another baseline vector within the triad is identical to the baseline vector $j$ (as is possible for linear triads) and $q_j=1$ otherwise, $k=(k_\parallel^2 + k_\perp^2)^{1/2}$ is the magnitude of the comoving wavenumber and the second equality uses the fact that $P(k)$ is invariant under the orientation of $(\mathbf{k}_\perp, k_\parallel)$ due to the isotropy of the cosmic signal. This expression shows that the closure phase delay power spectrum is proportional to an average of $P(k)$ weighted by the window function. In particular, each baseline of a closure triad maps to its own $\mathbf{k}_\perp$-mode, as is represented by the sum over baselines in the expression of the window function. This means that there is no exact one-to-one mapping between non-equilateral closure triads and $k_\perp$-space. The topography of the window function is determined further by the bandwidth, the spectral tapering, and the antenna primary beam. As a result, power spectrum estimates are always affected by some amount of 'mode-mixing', even if they are derived from equilateral triad measurements or from visibilities. If we disregard the more subtle effects of mode-mixing, we can indeed associate an equilateral triad with a unique $|k_\perp|$-mode since there is only one baseline length involved. For compact arrays such as HERA, the finite bandwidth often dominates the mode-mixing (to be discussed in Section~\ref{sec:mode-mixing}), even when a triad comprises baselines of different lengths. Moreover, the window function is often sufficiently localised around some value of $k$ for $P(k)$ to be approximated as constant across the window. In such cases, we can take $P(k)$ out of the integral and estimate the cosmological power spectrum as

\begin{align}
    \label{eq:ps_est}
    \widehat{P}_\triangledown(\tau) &= 2 \left(\frac{\lambda^2}{2k_\mathrm{B}}\right)^2 \left(\frac{1}{\Omega_\triangledown} \right) \left|\widetilde{\Psi}_\triangledown^\mathrm{H}(\tau)\right|^2.
\end{align}
The quantity $\Omega_\triangledown$ is the volume of the window function and can be written as
\begin{align}
    \label{eq:omega_closure}
    \nonumber \Omega_\triangledown &= \smashoperator{\int\limits_{-\infty}^{+\infty}} \frac{d^2k_\perp dk_\parallel}{(2\pi)^3} \sum_{j=1}^{3} q_j \left|\widetilde{\Xi}_j^{}\left(\tau - \frac{X k_\parallel}{2\pi}\right)\widetilde{A}\left(\mathbf{b}_{j} - \frac{Y\mathbf{k}_\perp}{2\pi}\right)\right|^2 \\ \nonumber 
    &= \frac{1}{XY^2}  \smashoperator{\int\limits_{-\infty}^{+\infty}} dldm \,\left|A\left(\mathbf{\hat{s}}\right)\right|^2 \sum_{j=1}^{3} q_j \smashoperator{\int\limits_{-\infty}^{+\infty}} d\nu^\prime \left|\frac{W(\nu^\prime)}{V_j^\mathrm{F}(\nu^\prime)}\right|^2,
\end{align}
where on the second line we split the integral into its transverse and line-of-sight components, changed the integration variables, and used Plancherel's theorem. The expression for the window function motivates the definitions of the beam solid angle, $\Omega_\mathrm{pp}$, the effective bandwidth, $B_\mathrm{eff}$, and the effective visibility, $V_\mathrm{eff}$, as
\begin{align}
    \Omega_\mathrm{pp} &= \smashoperator{\int\limits_{-\infty}^{+\infty}} dldm \, \left|A(\hat{\mathbf{s}}, \nu)\right|^2, \\ 
    B_\mathrm{eff} &= \smashoperator{\int\limits_{-\infty}^{+\infty}} d \nu \left|W\left(\nu \right)\right|^2, \quad \mathrm{and} \\
    \label{eq:veff}
    V_\mathrm{eff}^{-2} &= \sum_{j=1}^{3} q_j \left(\hat{V}_j^\mathrm{F}\right)^{-2},
\end{align}
respectively, where
\begin{equation}
    \label{eq:v_hat}
    \left(\hat{V}_j^\mathrm{F}\right)^{-2} = \frac{1}{B_\mathrm{eff}} \smashoperator{\int\limits_{-\infty}^{+\infty}} d\nu \left|\frac{W(\nu)}{V_j^\mathrm{F}(\nu)}\right|^2.
\end{equation}
Using these expressions and writing out the scaling parameters X and Y, we can write Equation~\ref{eq:ps_est} as
\begin{equation}
    \label{eq:ps_estimate}
     \widehat{P}_\triangledown(\tau) = 2\left(\frac{\lambda^2}{2k_\mathrm{B}}\right)^2 \left(\frac{c (1+z)^2 D_c^2 }{H(z) \nu_\textsc{H\,i}}\right) \left(\frac{V_\mathrm{eff}^2}{\Omega_\mathrm{pp} B_\mathrm{eff}} \right) \left|\widetilde{\Psi}_\triangledown^\mathrm{H}(\tau)\right|^2.
\end{equation}

\begin{figure}
    \centering
    \includegraphics[width=\linewidth]{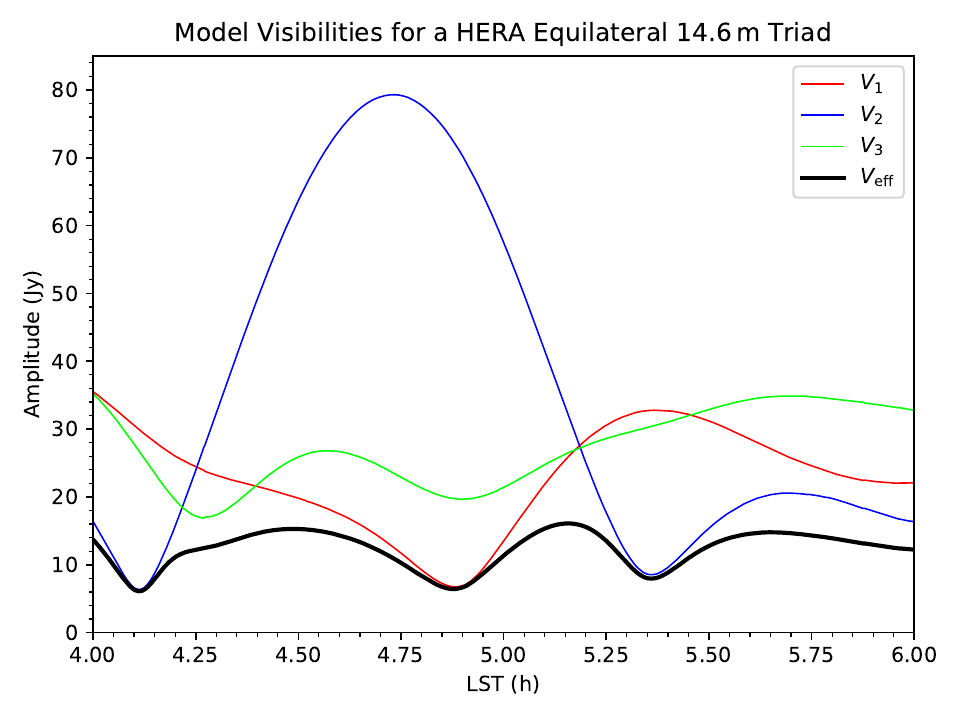}
    
    \caption[Effective visibility]{Model visibility amplitudes for a HERA equilateral 14.6\,m triad (red, blue and green) as a function of local sidereal time (LST). The thick black line represents the effective visibility, $V_\mathrm{eff}$, which provides the scaling of the closure phase delay power spectrum. This example neglects the frequency dependence of the visibility amplitudes.}
    \label{fig:veff_model}
\end{figure}

The normalisation in the above expression differs from that of the standard visibility delay power spectrum in that it contains an additional factor of two due to the fact that only half of the \textsc{H\,i} fluctuations are recovered in the visibility phases (see Table~\ref{tab:comparisson}).\footnote{\cite{closurelimits} and \cite{Keller2023} include an additional factor of 1/3 in the power spectrum scaling to account for the contribution of three visibility phases to the closure phase. This would be justified if $V_\mathrm{eff}$ was defined by an average rather than a sum of $\hat{V}_j^\mathrm{F}$ over baselines in inverse quadrature. Since this is not how \cite{closurelimits} define $V_\mathrm{eff}$, it appears that the additional factor of 1/3 stems from a bookkeeping mistake.} The effective visibility can be thought of as that part of $\Omega_\triangledown$ which differs from the conventional scaling of the visibility-based power spectrum estimate. While providing the estimated power spectrum with the correct units, $V_\mathrm{eff}$ also counteracts the inverse proportionality of the closure phase perturbations to the foreground visibility (see Equation~\ref{eq:cp_pert}). Since $V_\mathrm{eff}$ is obtained by averaging $V_j^\mathrm{F}(\nu)$ in inverse quadrature, it tends towards the lowest visibility amplitude within the closure triad, where perturbations to the visibility phase are the strongest (see Equation~\ref{eq:phase_taylor} and Figure~\ref{fig:veff_model}). In practice, $V_j^\mathrm{F}(\nu)$, and hence $V_\mathrm{eff}$, can be obtained from calibrated or modelled data. The requirements on the fractional accuracy of the calibration solutions or the model are not as high as in standard approaches because the sole purpose of estimating $V_j^\mathrm{F}(\nu)$ is to obtain the power spectrum scaling and possibly to compute window functions, but not to correct the data in any other way that could impart spectral structure on the data \citep[cf.][]{closuremaths}. 

\begin{table*}[t]
    \centering
    \renewcommand{\arraystretch}{3}
    \caption{Comparison of terms used in different delay spectrum approaches.}
    \resizebox{\textwidth}{!}{
    \begin{tabular}{c | c c c}
        \toprule\hline
        Term & Visibility & Closure phase prev. & This work \\
        \midrule
        $\widehat{P}(\tau)$ 
            & $\propto \left|\widetilde{V}^\mathrm{H}(\tau)\right|^2$ 
            & $\propto \frac{2}{3} V_\mathrm{eff} \left|\widetilde{\Psi}_\triangledown^\mathrm{H}(\tau)\right|^2$ 
            & $\propto 2 V_\mathrm{eff} \left|\widetilde{\Psi}_\triangledown^\mathrm{H}(\tau)\right|^2$ \\
        $\left(\hat{V}_j^\mathrm{F}\right)^{-2}$ 
            & -- 
            & $\left(\frac{\int d\nu W(\nu) V_j^\mathrm{F}(\nu)}{\int d\nu W(\nu)}\right)^{-2}$ 
            & $\frac{1}{B_\mathrm{eff}} \int d\nu \left|\frac{W(\nu)}{V_j^\mathrm{F}(\nu)}\right|^2$ \\
        $W(\mathbf{k_\perp}, k_\parallel; \tau)$ 
            & $\left<\left|\widetilde{W}\left(\tau - \frac{X k_\parallel}{2\pi}\right)\widetilde{A}\left(\mathbf{b} - \frac{Y \mathbf{k}_\perp}{2\pi}\right)\right|^2\right>$ 
            & $\sum_{j=1}^{3} q_j \left<\left|\widetilde{\Xi}_j\left(\tau - \frac{X k_\parallel}{2\pi}\right)\widetilde{A}\left(\mathbf{b}_{j} - \frac{Y \mathbf{k}_\perp}{2\pi}\right)\right|^2\right>$ 
            & $\sum_{j=1}^{3} q_j \left<\left|\widetilde{\Xi}_j\left(\tau - \frac{X k_\parallel}{2\pi}\right)\widetilde{A}\left(\mathbf{b}_{j} - \frac{Y \mathbf{k}_\perp}{2\pi}\right)\right|^2\right>$ \\\hline
        \bottomrule
    \end{tabular}
    }
    \tablefoot{This table compares terms used in the standard delay spectrum approach (first column), the closure phase approach according to \cite{closuremaths} (second column) and that according to this work (third column). From top to bottom, the rows show the terms used to estimate the power spectrum, the visibility-like term used to compute the effective visibility (cf. Equation~\ref{eq:veff}; relevant only to the closure phase approaches), and the cylindrical window function. For the power spectrum terms, we have omitted parts that are common to all three approaches and use $\widetilde{V}^\mathrm{H}(\tau)$ to denote the standard visibility delay spectrum of the cosmic \textsc{H\,i} signal. In the zeroth ('Term') column, we have omitted the $\triangledown$ subscript to make the terms representative of both the visibility and the closure phase approaches. See the text for the definitions of the other symbols used in this table.}
    \label{tab:comparisson}
\end{table*}

\subsection{Comparison with previous work}
The effective visibility was first introduced by \cite{closuremaths} as a heuristic amplitude scaling to gauge the cosmic \textsc{H\,i} signal fluctuations recovered in the closure phase to those in visibilities. However, they estimated $\hat{V}_j^\mathrm{F}$ as a windowed average of $V_j^\mathrm{F}(\nu)$ over frequency (see Table~\ref{tab:comparisson}). In contrast, our derivation shows that $\hat{V}_j^\mathrm{F}$ should be obtained by averaging $|V_j^\mathrm{F}(\nu)|$ in inverse quadrature across the frequency band (Equation~\ref{eq:v_hat}). This form of $V_\mathrm{eff}$ arises naturally from the normalisation of the effective delay window (Equation~\ref{eq:xi}) of the closure phase delay spectrum. The difference between the two definitions of $V_\mathrm{eff}$ should be subtle if $V_j^\mathrm{F}(\nu)$ varies slowly and smoothly with frequency, but could otherwise lead to substantial biases in the estimated power spectrum. The estimator presented here is free of such biases, provided that the assumptions of the derivation are met, i.e. $V_j^\mathrm{F}(\nu) \gg V_j^\mathrm{H}(\nu)$, and the correlations $C(\mathbf{b}_j, \mathbf{b}_{j^\prime}, \tau, \tau^\prime)$ are small between non-identical baselines in a closure triad.

\subsection{Mode-mixing}
\label{sec:mode-mixing}
In deriving $\widehat{P}_\triangledown(\tau)$, we assumed that the true cosmological power spectrum is approximately constant within the region probed by the window function. Generally, this is not necessarily the case, but the scaling of $\widehat{P}_\triangledown(\tau)$ nevertheless accounts for the correct normalisation of the window function. The window function provides information on how different power spectrum modes contribute to $\widehat{P}_\triangledown(\tau)$ and is therefore essential for the correct interpretation of the estimated power spectrum. 

The mode-mixing in a closure phase based analysis differs from that in a visibility-based analysis in two particular ways (see Table~\ref{tab:comparisson}). Firstly, the line-of-sight window function depends on the inverse foreground visibilities and the foreground closure phase through the term $\Xi_j(\nu) = \exp\{i\phi_\triangledown^\mathrm{F}(\nu)\} / V_j^\mathrm{F}(\nu)$ (cf. Equation~\ref{eq:xi}). This additional spectral component broadens the line-of-sight window function, resulting in a stronger mixing of neighbouring $k_\parallel$-modes. Secondly, $\widehat{P}_\triangledown(\tau)$ receives contributions from up to three distinct baselines, which results in additional $k_\perp$ mode-mixing. In order to associate a given triad and delay mode with a unique $k$-value, we propose computing a representative $k_\perp$ by defining an effective baseline length in place of $\mathbf{b}_\nu$ (cf. Equation~\ref{eq:veff}):

\begin{equation}
    b_\mathrm{eff} = V_\mathrm{eff}^{2}\sum_{j=1}^{3} b_j q_j \left(\hat{V}_j^\mathrm{F}\right)^{-2}.
\end{equation}
This definition is motivated by the fact that the contribution of each baseline to the closure phase delay spectrum is inversely proportional to the foreground visibility amplitude (see Equation~\ref{eq:phase_taylor}). For HERA baselines, this approach is sufficient, since we are generally in the regime where $k_\perp \ll k_\parallel$ and hence $k \sim k_\parallel$. Henceforth, we use $b_\mathrm{eff}$ to compute unique $k$-values for any given triad and delay mode, as this allows us to plot closure phase power spectra as a function of $k$ and make direct comparisons with standard power spectra. Note that $\widehat{P}_\triangledown(k)$ does not correspond exactly one-to-one to the cosmological power spectrum, but may approximate it to high accuracy if the window function is sufficiently localised. We investigate the accuracy of using $\widehat{P}_\triangledown(k)$ as an estimator of the cosmological power spectrum in Section~\ref{sec:results}.

Finally, it is instructive to determine observational parameters at which the mixing of $k_\parallel$-modes and that of $k_\perp$-modes take on an equal extent. Consider a typical bandwidth of $10\,$MHz and a central frequency of $165\,$MHz ($z=7.9$) as used in previous HERA studies \citep{upperlimits, upperlimits2, closurelimits, Keller2023}. The resolution of $k_\parallel$ is then at most $0.056\,h\,$Mpc$^{-1}$ (see Equation~\ref{eq:k}), but may be coarser depending on the spectral tapering used and the spectral properties of the foregrounds. Taking the same number for the resolution of $k_\perp$ and translating it to a difference in baseline lengths, we find a value of $\sim 100$\,m. In this example, bandwidth is the dominant source of mode-mixing for baseline length differences smaller than $100$\,m. For longer baseline length differences $k_\perp$-mixing starts to dominate.

\subsection{Foregrounds}
Astrophysical foregrounds have smooth spectra at frequencies of interest to cosmological \textsc{H\,i} 21\,cm studies. Their imprint on the delay spectrum is therefore expected to be confined to low delay modes. However, the chromatic response of an interferometer makes the foreground continuum appear unsmooth, which broadens the foreground dominated region in delay space. For flat-spectrum foregrounds and in the absence of instrumental systematic effects, foreground power in the standard visibility delay spectrum is bounded by the horizon limit, defined as $\tau_\mathrm{h,j} = |\mathbf{b}_j| / c$ \citep{dspec, eorwindowmath}. The foreground contaminated region of a delay spectrum is termed the 'wedge' due to its wedge-shaped appearance in cylindrical $(k_\perp, k_\parallel)$-space \citep[][]{datta}.

\cite{closuremaths} find two additional effects that broaden the foreground component in the closure phase delay spectrum. Firstly, the foreground phase term, $\exp\{i\phi_\triangledown^\mathrm{F}\}$, is a triple product of three individual visibility phase terms, resulting in a triple convolution in delay space. This effect proliferates foreground power up to $\tau_{\triangledown, \mathrm{h}} = \sum_j |\mathbf{b}_j| / c$. Secondly, higher-order terms in the closure phase approximation cause foreground power to leak up to arbitrarily high delay modes. These higher order terms are typically small enough to be neglected \citep{closuremaths}. In this work, we assume delay modes $|\tau| > \tau_{\triangledown, \mathrm{h}} + \tau_\mathrm{b}$ to be uncontaminated by foregrounds, where $\tau_\mathrm{b}=0.39\,\mu$s is a buffer to account for other spectral effects such as foreground and beam chromaticity (see Section~\ref{sec:beam_freq}), and the effects of computing the delay spectrum over a finite bandwidth. This value ensures a dynamic range of $10^{10}$ outside the buffer region and was empirically determined by investigating the delay window functions obtained from simulated data (to be discussed in Section~\ref{sec:validation}). The value of $\tau_\mathrm{b}$ is ultimately dependent on the effective bandwidth used to compute the delay spectrum \citep[cf.][]{Thyagarajan2016}.

\subsection{Beam chromaticity}
\label{sec:beam_freq}
So far, we have assumed the frequency dependence of the primary beam to be negligible (cf. Equation~\ref{eq:dvis_corr}). This is a commonly made assumption and usually approximates the primary beam well enough for a simple power spectrum estimation. However, in principle, there is no need to make this approximation \citep[see][]{Thyagarajan2016}. This leads to a modification of $W_\triangledown(\mathbf{k_\perp}, k_\parallel; \tau)$, where the delay transformed primary beam is convolved with $\widetilde{\Xi}_j(\tau)$. The volume of the new window function is then

\begin{equation}
    \Omega_\triangledown^\prime = \frac{1}{XY^2} \, \sum_{j=1}^{3} q_j \smashoperator{\int\limits_{-\infty}^{+\infty}} dldmd\nu^\prime \left|A\left(\mathbf{\hat{s}}, \nu\right)\frac{W(\nu^\prime)}{V_j^\mathrm{F}(\nu^\prime)}\right|^2.
\end{equation}
In this expression, the terms due to the beam, the bandwidth, and the foregrounds can no longer be separated into a product of $\Omega_\mathrm{pp}^{-1}$, $B_\mathrm{eff}^{-1}$, and $V_\mathrm{eff}^2$.
In this work, we use the achromatic approximation, as we find that the relative difference between $\Omega_\triangledown^\prime$ and $\Omega_\triangledown$ is only $\sim 10^{-3}$, which is at least an order of magnitude below our targeted accuracy. Moreover, the approximation is computationally more efficient and better aligned with the previous closure phase studies.

It is also worth noting that the frequency-dependence of the antenna beams can add spectral structure to the otherwise smooth foregrounds, especially when they lie in the sidelobes of the primary beam. This effect has been investigated by \cite{Thyagarajan2016} for the standard visibility approach and by \cite{charles2022simulations} for the closure phase approach. Both studies find that beam chromaticity induces foreground leakage beyond the wedge and report values of $0.2-0.3\,h\,$Mpc$^{-1}$, below which the leakage may preclude the detection of the cosmological signal.

\section{Validation}
\label{sec:validation}
\begin{figure}
    \centering
    \includegraphics[width=\linewidth]{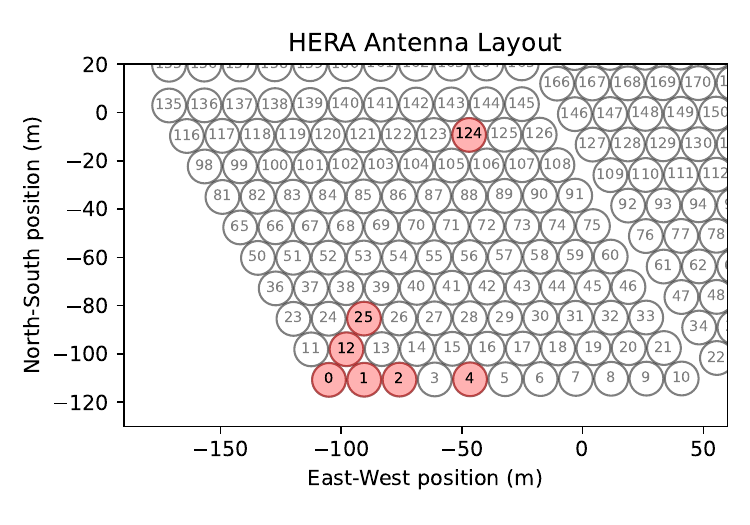}
    
    \caption[Array Layout]{HERA antennas used for validation (red). We only used triads that include antenna 0 to compute a complete and independent set of closure phases.}
    \label{fig:array}
\end{figure}

To validate our estimator, we simulated visibilities for 7 HERA antennas using the \texttt{matvis}\footnote{\url{https://github.com/HERA-Team/matvis}} simulator module through \texttt{hera\_sim}\footnote{\url{https://hera-sim.readthedocs.io/en/latest/}} \citep{matvis}. The antennas were chosen to include a variety of triad shapes, such as equilateral, linear, and scalene triads. The baseline lengths range from 14.6\,m to 116.8\,m, allowing us to investigate the mixing of different transverse scales in triads with extreme baseline length ratios. In total, there are 15 triads, which form a complete and independent set of closure phases \citep[c.f.][]{TMS2017}. Here, we used only triads that include antenna 0 (see Figure~\ref{fig:array}).

\subsection{Sky models}
Our simulations used sky models consisting of foreground sources and a cosmic line signal, weighted by a frequency-dependent model of the HERA primary beam \citep{beam}. For the foreground model, we selected point sources from the GLEAM catalogue \citep{gleam} with flux densities greater than 50 mJy at 200\,MHz. For the cosmic line signal, we used two different models: 

\begin{itemize}
    \item Model 1: Gaussian random field with ionised regions.\\
    This is a toy model of the EoR that starts with a Gaussian random temperature field, $T(l, m)$, with a power-law spectrum $\sim k^{-\alpha/2}$, mean $\delta T_0$, and standard deviation $\sigma_T$. We then add 'ionised regions' by setting the temperature field to zero where $T(l, m) > T_c$. Here, we chose $\alpha=2$, $\sigma_T=3\,$mK and $T_c=\delta T_0=10\,$mK, which results in a temperature field that is approximately half ionised. The advantage of this simple toy model is that it allows us to efficiently generate a large number of independent samples. We computed 300 independent realisations of this model spanning $4.8\,$Gpc and $3072$ cells along each transverse axis, thus covering around 30 square degrees of the sky at the redshifts of the EoR. \\

    \item Model 2: \texttt{21cmFast} light cones.\\
    This is a more realistic representation of the \textsc{H\,i} 21\,cm reionisation signal generated with the \texttt{21cmFast}\footnote{\url{https://21cmfast.readthedocs.io/en/latest/}} simulator \citep{mesinger2011, murray2020}. We ran \texttt{21cmFast} with an ionising efficiency $\zeta=20$, a minimum virial temperature of star forming haloes $T_\mathrm{vir}=10^4$~K, and left the other astrophysical parameters at their default setting. In our simulations, we did not account for spin temperature fluctuations. Due to the computational cost of these simulations, we generated only 12 light cones with volumes of $1.6^3\,$Gpc$^3\,$ and $1024^3$ cells. At EoR redshifts these light cones cover only around 10 square degrees of the sky. This is not enough to fill the primary beam of a HERA antenna \citep[FWHM $\sim 10^\circ$;][]{beam}. For this reason, we followed the same procedure as in \cite{closurelimits} and \cite{Keller2023} and tiled the light cones three times along each transverse axis. Each tile is a reflected version of its neighbouring tiles to avoid discontinuities at the tiling boundaries. 
\end{itemize}
Following the same procedure as in \cite{closurelimits} and \cite{Keller2023}, we smoothed and downsampled the above models to an angular resolution of $\sim 7^\prime$, resulting in 256 pixels per transverse axis. This downsampling is appropriate for the $\sim 0.5^\circ$ FWHM of the HERA synthesised beam and reduces the computational cost of the subsequent visibility computation.

\subsection{Power spectrum computation}
We simulated visibilities for the foregrounds visible to HERA between 0.0\,h < RA < 6.25\,h and 21.5\,h < RA < 24.0\,h, corresponding to sky regions that are devoid of bright galactic diffuse emissions. We performed a direct Fourier transform on the beam weighted EoR cubes to obtain visibilities which we then added to a uniformly selected sample of foreground visibilities.

We computed delay spectra using a standard fast Fourier transform (FFT) with a modified Blackman-Harris window as our spectral tapering function \citep{blackman-harris, Thyagarajan2016}. The modified Blackman-Harris window is a regular Blackman-Harris function convolved with itself for increased spectral dynamic range. The drawback of the modified spectral tapering function is that it has a broader response in delay space \citep[noise equivalent width: $w_\mathrm{ne} \approx$ 2.88;][]{Thyagarajan2016}. For this reason, we downsampled the delay spectra by a factor of 3 along the delay axis to contain only samples that are approximately independent. The band over which we computed delay spectra is $152 - 167$\,MHz, corresponding roughly to Band 2 used in \cite{upperlimits2} to set upper limits on the EoR \textsc{H\,i} 21\,cm signal at redshift $z=7.9$. We then computed power spectra according to Equation~\ref{eq:ps_estimate}, where we used the cosmological parameters from \cite{planck2018param} to compute the scaling and obtained $V_\mathrm{eff}$ directly from the simulated foreground visibilities, and $\Omega_\mathrm{pp} \approx 0.016$ from the HERA primary beam model \citep{beam}. 

When averaging power spectra, we flagged instances where $V_\mathrm{eff} < 1\,$Jy to ensure that the assumptions of the closure phase approximation are met (cf. Section~\ref{sec:pert}). In practice, when estimating $V_\mathrm{eff}$ from calibrated data, it is desirable to have enough signal-to-noise ratio in the visibility amplitudes to avoid biases in the power spectrum scaling. Moreover, as discussed in the next section, low visibility amplitudes can lead to a wider delay response and increased foreground leakage. The flagging on $V_\mathrm{eff}$ helps avoid such instances.

Alongside power spectra, we also computed the 'C-terms' (Equation~\ref{eq:cp_dspec_corr}) and window functions (i.e. Equation~\ref{eq:w_cyl} and the components thereof) using FFTs. 

\section{Results}
\label{sec:results}
This section presents the results of our validation efforts. We discuss the window functions first, as these offer the necessary information about the amount of mode-mixing in a given measurement to decide whether to include it in further analysis. We then test the accuracy of our power spectrum estimator against the standard visibility estimator and compare them with the intrinsic power spectra of our models. Lastly, we discuss the effect of inter-mode C-terms in our estimator.

\subsection{Delay window functions}
\label{sec:delay_window}
We investigate the effective delay window function, $\widetilde{\Xi}(\tau) = \sum_j q_j \widetilde{\Xi}_j(\tau)$, which determines the amount of line-of-sight mode-mixing in the closure phase delay spectrum. Unlike in the standard visibility delay spectrum, this delay window function is inherently foreground dependent. Spectral structure in the foreground visibility phases can lead to intolerable amounts of line-of-sight mode-mixing and foreground leakage beyond the horizon limit. This is especially true for low visibility amplitudes, where small spectral variations in the visibilities can have a strong effect on $\Xi_j(\nu)$. The amount of spectral structure in $\Xi_j(\nu)$ therefore depends on the baseline length and orientation, as well as the apparent sky.

\begin{figure}
    \centering
    \includegraphics[width=0.93\linewidth]{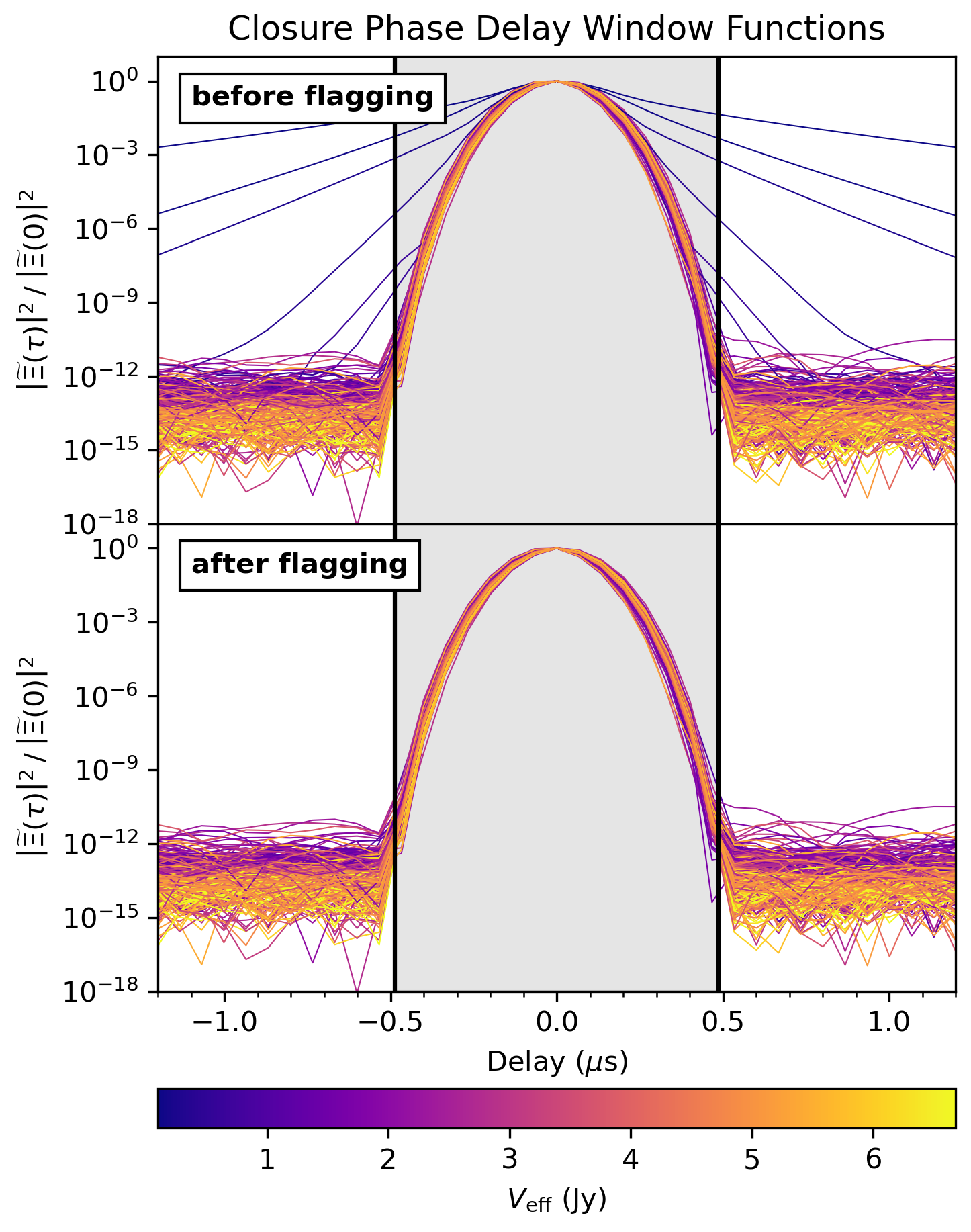}
    \caption[Delay Windows]{Normalised closure phase delay window functions for an equilateral 14.6\,m triad before (top) and after (bottom) flagging instances with low dynamic range ($<10^{10}$). The colour indicates the corresponding value of the effective visibility, showing that low dynamic range windows tend to be associated with low visibility amplitudes and hence low values of $V_\mathrm{eff}$. The shaded region indicates delays which lay within the horizon limit of the closure phase delay window function plus buffer.}
    \label{fig:delay_window}
\end{figure}

Figure~\ref{fig:delay_window} shows $|\widetilde{\Xi}(\tau)|^2$ for an equilateral 14.6\,m triad and 300 different realisations of the foreground sky, colourised by the corresponding values of $V_\mathrm{eff}$. As expected, some of the window functions are severely limited in dynamic range. Furthermore, the plot confirms our intuition about the scaling of the dynamic range with $V_\mathrm{eff}$. We propose flagging instances where a certain dynamic range is not achieved within a given delay range. We determine a suitable delay range by considering the expected width of the window function. Since $\Xi_j(\nu)$ consists of two visibility phase terms ($\phi_j$ cancels out in the division by $V_j^\mathrm{F}$), its maximum delay is approximately $\tau_{\mathrm{wh}, j} = \tau_{\triangledown,\mathrm{h}} - \tau_{\mathrm{h}, j}$ in the limit of spectrally flat visibility amplitudes. Hence, we define the horizon limit of the closure phase delay window function simply as $\tau_\mathrm{wh} = \max(\{\tau_{\mathrm{wh}, 1}, \tau_{\mathrm{wh}, 2}, \tau_{\mathrm{wh}, 3})$. Motivated by the fact that foregrounds overpower the EoR \textsc{H\,i} 21\,cm line by a factor of $\sim 10^5$, we flag instances where the dynamic range of the squared window function is smaller than $10^{10}$ at delays $|\tau| < \tau_\mathrm{wh} + \tau_\mathrm{b}$ (see lower graph in Figure~\ref{fig:delay_window}). For our foreground model, the flagging fractions range from 4\% to 30\%. The flagging fraction tends to increase with the lengths of the triad baselines due to their increased chromaticity, which tends to broaden the delay window function and limit the spectral dynamic range (see Section~\ref{sec:mode-mixing}). 

\subsection{Power spectra}
\begin{figure*}
    \centering
    \includegraphics[width=0.97\linewidth]{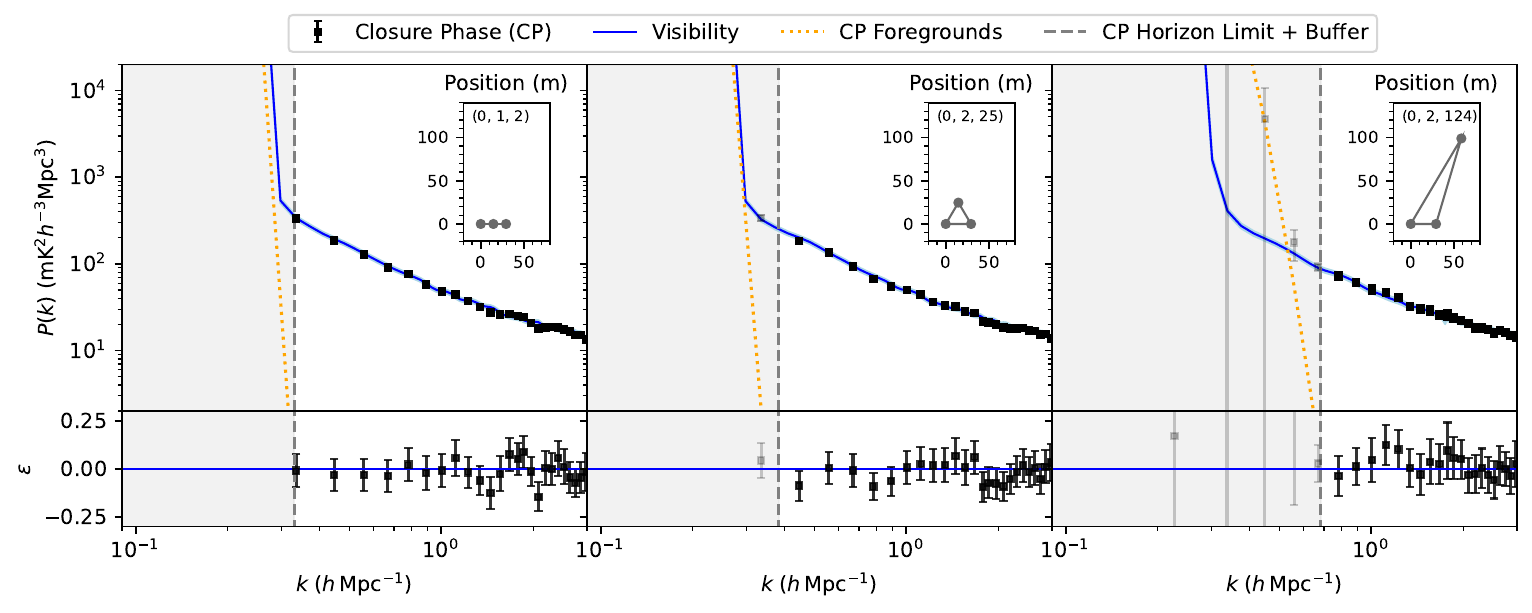}
    \caption[Toy Model Power Spectra]{Model~1 power spectrum estimates for three triads (see inserts). The top panel shows power spectra estimated from visibilities (solid blue line) and closure phases (black scatter points). The dotted orange line shows the closure phase power spectrum for a sky consisting of foregrounds only. The lower panel shows the fractional difference, $\varepsilon$, between the visibility and closure phase power spectra. The shaded regions mark the foreground wedge. All uncertainties due to cosmic variance are $2\sigma$.}
    \label{fig:toy_ps}
\end{figure*}

Figure~\ref{fig:toy_ps} shows the averaged visibility and closure phase power spectra of Model~1 for three different triads: a linear east-west triad with two 14.6\,m baselines, an equilateral 29.2\,m triad, and a scalene triad with one short baseline (29.2\,m) and two long baselines (105.3\,m and 116.8\,m). The flagging fraction was 6.6\%, 7\%, and 30\%, respectively (see previous section). The three visibility power spectra of a triad were averaged in bins of $k$ to obtain the plotted visibility power spectrum (solid blue line). The closure phase delay power spectrum of the foregrounds is also shown (dotted orange line). All uncertainties in the plots are $2\sigma$.

The broadening of the foreground wedge with increasing triad size is clearly discernible; while the wedge extends to $k \sim 0.3-0.4\,h\,$Mpc$^{-1}$ for the linear and equilateral triads, it extends to $k \sim 0.7\,h\,$Mpc$^{-1}$ for the long-baseline triad. For the latter, the increased size of the foreground region of the closure phase power spectrum with respect to that of the visibilities is particularly evident. 

As can be seen in the residual plots (lower panel of Figure~\ref{fig:toy_ps}), the closure phase and visibility power spectra agree well within the $2\sigma$ uncertainty. The RMS values of the fractional residuals are 3.3\%, 4.2\% and 5.3\%, respectively, and can be expected to decrease further as more samples are averaged.

\begin{figure*}
    \centering
    \includegraphics[width=0.45\linewidth]{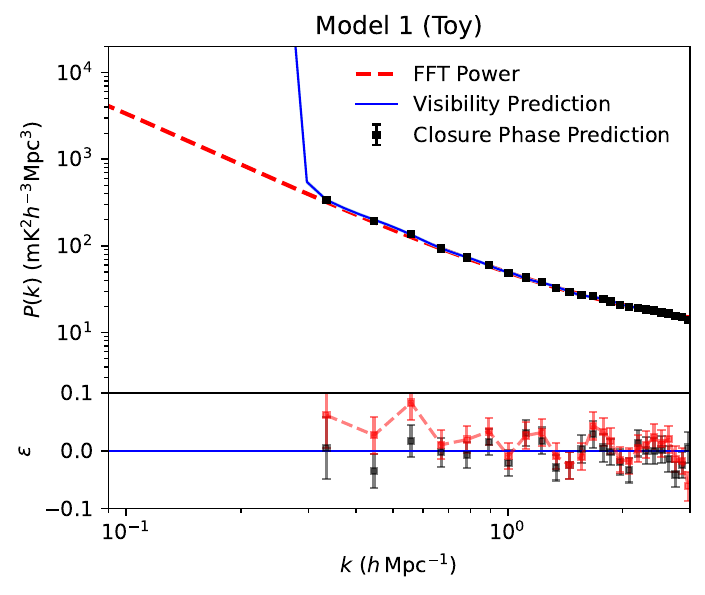}
    \includegraphics[width=0.45\linewidth]{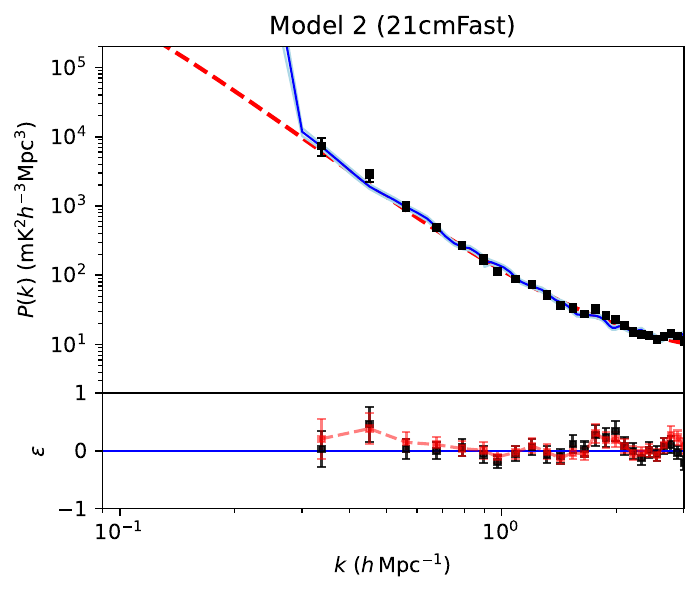}
    \caption[Averaged Power Spectra]{Same as Figure~\ref{fig:toy_ps} but averaged across all 15 triads in bins of $k$. The left and right plots show the estimated power spectra for Model~1 and Model~2, respectively. The red dashed line shows the power spectrum obtained from doing a 3D FFT on the full resolution cubes.}
    \label{fig:ps_avg}
\end{figure*}

We averaged the power spectra of all 15 triads to further test the accuracy of our power spectrum estimator. Figure~\ref{fig:ps_avg} shows averaged power spectrum estimates for Model 1 (left) and Model 2 (right), respectively, alongside a more accurate estimate of the intrinsic power obtained from a 3D FFT of the full-resolution model cubes (dotted red lines). The visibility and closure phase results continue to agree with each other at high significance. While there are a few $2\sigma$-outliers in the residual plots of both models, they are still consistent with zero at the $3\sigma$-level. Note that the residuals are strongly correlated, so neighbouring modes of an outlier point are more likely to be outliers. Such accumulation of outliers can be seen, for example, in the right plot of Figure~\ref{fig:ps_avg} at around $k \sim 2\,h\,$Mpc$^{-1}$. The overall consistency between the visibility and closure phase power spectra suggests that upper limits derived from the two approaches can be directly compared to each other once the relevant scaling factors have been accounted for.

The closure phase delay power spectra accurately estimate the intrinsic power at intermediate and high $k$-modes, but slightly overestimate it at $k \lesssim 0.6\,h\,$Mpc$^{-1}$ for Model~1. This is probably attributed to the smearing of power by the window function, as has already been noted in similar tests done by \cite{HERAvalidation} and has been discussed in Section~\ref{sec:mode-mixing} of this paper.

Although thermal noise was not included in our simulations, we note that the flagging fraction due to low $V_\mathrm{eff}$ would result in moderate sensitivity detriments of approximately 3\%, 4\%, and 16\%, respectively, assuming uniform sensitivity across the LST-range. In a real data reduction, this would be additional to the commonly applied flags of a visibility-based analysis (e.g. RFI and instrumental flags), although some overlap may be possible. The loss in sensitivity is relatively low, as most of the sensitivity comes from coherent averaging over consecutive nights and redundant baseline measurements \citep[see][]{Parsons2012sensitivity}. We leave an in depth discussion of the effects of thermal noise on the sensitivity of the closure phase power spectrum to future work.

\subsection{Cylindrical mode-mixing}
\begin{figure*}
    \centering
    \includegraphics[width=0.94\linewidth]{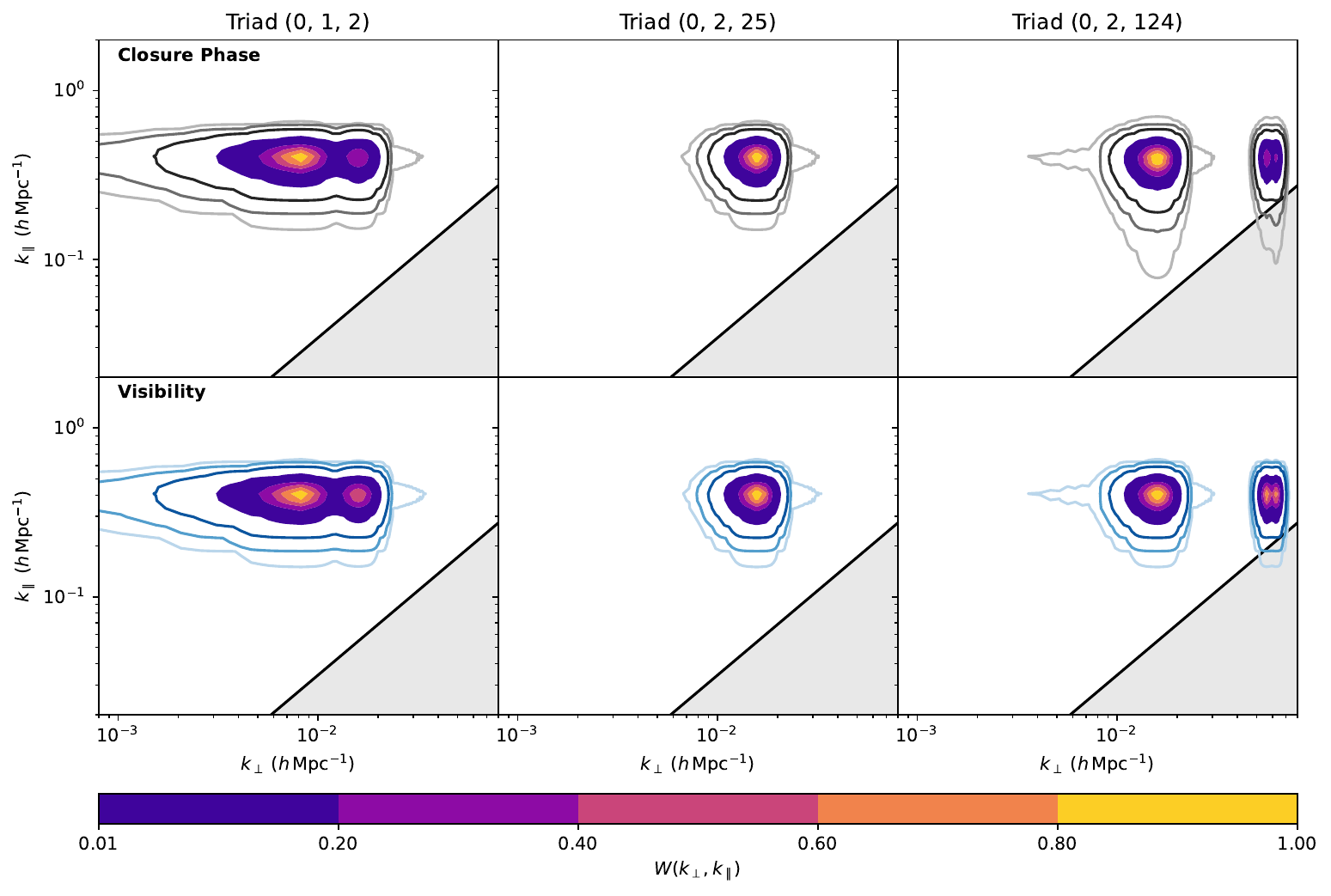}
    \caption[Cylindrical Windows]{Normalised cylindrical window functions at $k_\parallel = 0.4\,h\,$Mpc$^{-1}$ for closure phases (top row) and visibilities (bottom row). Each column shows the window functions for a different triad: a linear east-west triad with two 14.6\,m baselines, an equilateral 29.2\,m triad, and a scalene triad with one short baseline (29.2\,m) and two long baselines (105.3\,m and 116.8\,m) (cf. Figure~\ref{fig:toy_ps}). For visualisation purposes, the bottom row shows the sum of the visibility window functions of the three triad baselines. The contour lines are at the levels of $10^{-4}$, $10^{-6}$, and $10^{-8}$. The shaded regions in the bottom right corners indicate the standard wedge region.}
    \label{fig:window_cyl}
\end{figure*}

We investigate the relative contributions of different $(k_\perp, k_\parallel)$-modes to the power spectrum estimates. Figure~\ref{fig:window_cyl} shows cylindrical window functions of the power spectra at $k_\parallel = 0.4\,h\,$Mpc$^{-1}$ estimated from closure phases (top row) and from visibilities (bottom row). The columns in the figure correspond to the three triads discussed in the previous section. For better visualisation, the visibility window functions have been averaged across the three baselines. 

As expected, the number of peaks in the cylindrical power spectra corresponds to the number of unique baseline lengths in a given triad. Each peak is located at the $(k_\perp, k_\parallel)$-mode mapped by one of the triad baselines and the delay mode (see Equation~\ref{eq:k}). The relative amplitudes of the peaks are not necessarily equal. For example, the peak corresponding to the 29.2\,m baseline of the linear triad is only $20-40\,\%$ of the peaks corresponding to the 14.6\,m baselines (top left plot in Figure~\ref{fig:window_cyl}). This is a consequence of the foreground dependence of the window functions: The contribution of each baseline is weighted by the squared inverse of the visibility amplitude (see Equations~\ref{eq:xi} and \ref{eq:w_cyl}). Apart from this weighting, the closure phase and visibility window functions are almost indistinguishable. The main discrepancies between the closure phase and visibility window functions are seen at modes that are close to the foreground wedge. In our example, this is the case for the scalene triad (right column), where the closure phase window function has a noticeably larger extent in the $k_\parallel$-direction, especially at levels below $10^{-6}$. The window function thus extends further into the foreground wedge, thereby increasing the amount of foreground power leakage beyond the horizon limit. This broadening of the window function close to the wedge is caused by the $\exp\{i\phi_\triangledown^\mathrm{F}(\nu)\}$ term in the line-of-sight component of the window function (see Equation~\ref{eq:xi} and discussion in Sections~\ref{sec:mode-mixing} and \ref{sec:delay_window}).

Note that for the baselines used in this work, the scale of mode-mixing along the line-of-sight dimension is nearly an order of magnitude greater than along the transverse dimension. The spherically averaged window functions (that is, averaged in bins of $k$) is therefore approximately the same for closure phases and visibilities, despite their different imprint along the transverse dimension of the cylindrical power spectrum. 

\subsection{C-terms}
\label{sec:c-terms}
\begin{figure}
    \centering
    \includegraphics[width=\linewidth]{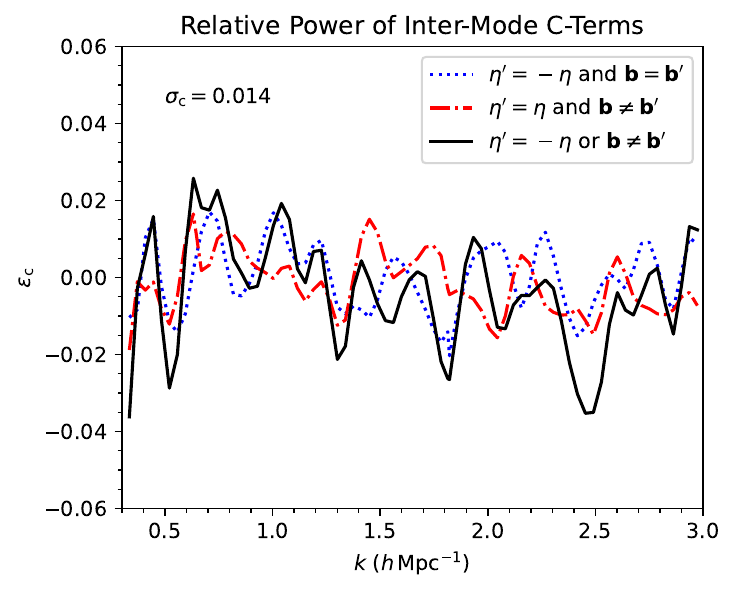}
    \caption[C-Terms]{Relative power of inter-mode C-terms in the averaged Model~1 power spectrum (left plot in Figure~\ref{fig:ps_avg}). The following C-terms are plotted: different delay modes but same baseline vector (blue dotted line), same delay modes but different baseline vectors (red dash-dotted line), and different delay modes and/or different baseline vectors (black line). The variance of the latter is $\sigma_\mathrm{c} = 0.014$.}
    \label{fig:c_terms}
\end{figure}

In our derivation of the power spectrum estimator, we assumed that inter-mode correlations are approximately zero (see Equation~\ref{eq:cp_dspec_corr_2}). Due to the finite sample size, these inter-mode C-terms constitute an additional noise-like component in the estimated power spectrum. We compute the relative contribution of these C-terms to the averaged Model~1 power spectrum (left plot in Figure~\ref{fig:ps_avg}) as

\begin{equation}
    \epsilon_\mathrm{c}(k) = \frac{\left<\sum_\mathrm{inter-modes} C(\mathbf{b}, \mathbf{b}^\prime, \tau, \tau^\prime)\right>_k}{\left<\sum_{j=1}^3 C(\mathbf{b}_j, \mathbf{b}_j, \tau, \tau) + C(\mathbf{b}_j, \mathbf{b}_j, -\tau, -\tau)\right>_k},
\end{equation}
where the sum in the numerator is over C-terms for which $\mathbf{b} \neq \mathbf{b}^\prime$ or $\tau = -\tau^\prime$, and the angular brackets denote an average over different sky regions and triads in bins of $k$ weighted by $V_\mathrm{eff}^2$ to correctly account for the relative contribution of each term to the final average. As can be seen in Figure~\ref{fig:c_terms}, $\epsilon_\mathrm{c}(k)$ appears noise-like and has a standard deviation of $\sigma_\mathrm{c} = 0.014$. We can compare this with the RMS of $\sim 0.018$ of the fractional residuals in the left plot of Figure~\ref{fig:ps_avg}, indicating that the C-terms account for approximately 60\% of the fractional difference between the closure phase and visibility power spectra. The remaining differences are mainly attributed to cosmic variance, which arises from the fact that the two estimates give different weights to different modes of the intrinsic power spectrum via their unique window functions.

For very compact triads, the inter-baseline C-terms should eventually stop averaging down like noise because the modes probed by the triad baselines are not entirely independent. As can be seen in Equation~\ref{eq:cp_dspec_corr_2}, the amount of additional signal recovered by inter-baseline C-terms depends on the primary beam and the difference between baseline vectors. Using the HERA beam model, we estimate the expected signal recovery in inter-baseline C-terms to be around 4\% for the shortest HERA baselines. The resulting bias can, in principle, be corrected for if higher accuracy is required. Here, we do not achieve a fractional accuracy high enough on individual triads to make this correction necessary. 

As the inter-mode C-terms are at most a few per cent of the cosmological signal strength, they are at least a factor of $10^{3}$ smaller than the thermal-like noise that limits current closure phase upper limits \citep[e.g.][]{Keller2023}. Future detections of the cosmological signal will have to quantify the contribution of inter-mode C-terms to the estimated power spectrum using a forward modelling approach as presented here.

\section{Summary and conclusion}
\label{sec:conclusion}
In this work, we have revised the closure phase delay spectrum technique for estimating the power spectrum of a cosmological line signal in the presence of strong foreground continuum emissions. Our estimator differs from previous works by our definition of the 'effective visibility'; a scaling factor which gauges spectral phase fluctuations to intensity fluctuations. As part of the derivation, we obtain the window function, which quantifies mode-mixing in our estimator. The effective visibility arises naturally as a part of the normalisation of the window function.

We have validated our estimator against the standard visibility delay spectrum technique using a subset of the HERA telescope and a sky model consisting of realistic point source foregrounds and two different models of the EoR \textsc{H\,i} 21\,cm signal. We find that the power spectrum estimates are consistent with those from visibilities within the uncertainty of cosmic variance, which is smaller than 2\% in our simulations. This indicates that the accuracy of the estimator may further increase as more samples are averaged, i.e. by including more triad shapes and orientations or more independent patches of the sky. 

The main differences between the closure phase and visibility delay spectrum techniques can be summarised as follows:

\begin{enumerate}
    \item \textit{Window functions}: The estimated power receives contributions from a range of intrinsic $k$-modes. This is due to the spatial boundaries of the measurement, set by the finite frequency bandwidth and the primary beam response. The mapping between the estimated and intrinsic power spectra is described by the window function. In contrast to the visibility estimator, the closure phase window function may peak at multiple $k_\perp$ if the triad baselines are of different lengths. Additionally, its line-of-sight component depends on the foreground visibilities, resulting in a broader response in $k_\parallel$-space. As a result, we find that low visibility amplitudes can severely reduce the width and dynamic range of the window function. However, after flagging such instances, we find that the width and shape of the closure phase window function are comparable to that resulting from the visibility-based technique.   

    \item \textit{C-terms}: Our power spectrum estimates are obtained by squaring and averaging the amplitudes of the closure phase delay spectra. In doing so, there are several inter-baseline and inter-mode cross-terms, which we call 'C-terms' (see Equation~\ref{eq:cp_dspec_corr_2}). We have shown that inter-mode C-terms average down to approximately zero and constitute an additional noise-like component in the closure phase power spectrum. 

    \item \textit{Higher-order terms}: The closure phase approximation uses linear perturbation theory to map spectral intensity fluctuations to phase fluctuations. \cite{closuremaths} have shown this approximation to reach a fractional accuracy of $10^{-4}$ for the expected strength of the EoR \textsc{H\,i} 21\,cm signal. This should be subdominant to the above effects and accurate enough for the initial detection and characterisation of the cosmic signal.

    \item \textit{Foreground wedge}: As already noted by \cite{closuremaths}, the foreground dominated region of the closure phase delay spectrum goes beyond the conventional foreground wedge. The broadening of the foreground wedge results from the multiplication of three visibility phase terms, which translates to a triple convolution in delay-space. The closure phase technique is therefore less sensitive to the lowest $k$-modes. This effect can potentially be mitigated by applying the delay transform directly to the unwrapped closure phase instead of its complex exponential \citep[See appendix of][]{closuremaths}.
\end{enumerate}

This work confirms the viability of estimating the cosmological \textsc{H\,i} 21\,cm power spectrum with closure phases. However, previous upper limits on the cosmic signal will have to be revised to account for the scaling differences. The improved understanding of mode-mixing will allow future studies to combine different triad shapes to achieve better sensitivity while accounting for the window functions. Our planned future work will focus on finding optimal ways to average the measurements of different closure triads in the presence of noise. Furthermore, we aim to improve the sensitivity of the approach, for example, by using other closure quantities that include visibility amplitude information \citep[e.g. closure invariants;][]{Thyagarajan+2022} and adapting techniques to further mitigate the foreground wedge \cite[cf.][]{closuremaths}. 

\begin{acknowledgements}
PMK is part of Allegro, the European ALMA Regional Centre node in the Netherlands. Allegro is funded by The Netherlands Organisation for Scientific Research (NWO/EW).

The software developed for this analysis uses Python and the publicly-accessible and open-sourced Python packages \texttt{21cmFast} \citep{mesinger2011, murray2020}, Numpy \citep{numpy}, SciPy \citep{scipy}, Astropy \citep{astropy}, Matplotlib \citep{matplotlib}, \texttt{matvis} and \texttt{hera\_sim} \citep{matvis}.
\end{acknowledgements}

\bibliographystyle{aa} 
\bibliography{references} 

\end{document}